\documentclass[doublecolumn,journal,a4paper,10pt]{IEEEtran}
\usepackage{flushend}
\usepackage{cite}
\usepackage{color}
\usepackage{epsf}
\usepackage{epsfig}
\usepackage{graphicx}
\usepackage{graphics}
\usepackage{epstopdf}
\usepackage[small]{caption}
\usepackage{amsmath}
\usepackage{amssymb}
\usepackage{amsthm}
\usepackage{amsxtra}
\usepackage[ruled]{algorithm2e}
\usepackage{algorithmic}
\usepackage{enumerate}
\usepackage{multirow}
\usepackage{enumerate}
\usepackage{amssymb}
\usepackage{lipsum}
\usepackage{setspace}
\usepackage{adjustbox}
\usepackage{multirow}
\usepackage{subcaption}
\usepackage{url}
\usepackage{color,soul}
\usepackage{geometry}
\usepackage[utf8]{inputenc}
\usepackage{array}
\usepackage{makecell}
\usepackage[normalem]{ulem}

\usepackage{rotating} % adde by Wayes Tushar
\usepackage{graphicx} % added by Wayes Tushar
\newcommand{\Rmnum}[1]{\expandafter\@slowromancap\romannumeral #1@}

\makeatother
\begin{document}
%\doublespace
%
%\title{Evaluation of Recent Technological Developments for Feasible Practice of Negawatt Trading}
\title{{Challenges and prospects for negawatt trading in light of recent technological developments}}
\author{Wayes Tushar$^{\text{a},*}$, Tapan K. Saha$^\text{a}$, Chau Yuen$^\text{b}$, David Smith$^\text{c}$, Peta Ashworth$^\text{a}$, H. Vincent Poor$^\text{d}$, and Subarna Basnet$^\text{e}$\\$^\text{a}$The University of Queensland, Brisbane, QLD 4072, Australia\\$^\text{b}$Singapore University of Technology and Design (SUTD), 8 Somapah Road, Singapore 487372\\$^\text{c}$CSIRO Data61, Eveleigh NSW 2015, Australia\\$^\text{d}$Princeton University, Princeton, NJ 08544, USA\\$^\text{e}$Massachusetts Institute of Technology, Cambridge, MA 02139, USA\\$^*$Corresponding author's e-mail: w.tushar@uq.edu.au}
\IEEEoverridecommandlockouts
\maketitle
%\thispagestyle{fancy}
%\thispagestyle{empty}
%\pagestyle{empty}
%\doublespace
% ================================
\begin{abstract}
With the advancement of the smart grid, the current energy system is moving towards a future where people can buy what they need, sell when they have excess, and can trade the right of buying to other proactive consumers (prosumers). While the first two schemes already exist in the market, selling the right of buying - also known as negawatt trading - is something that is yet to be implemented. Here, we review the challenges and prospects of negawatt trading in light of recent technological advancements. Through reviewing a number of emerging technologies, we show that the necessary methodologies that are needed to establish negawatt trading as a feasible energy management scheme in the smart grid are already available. Grid interactive buildings and distributed ledger technologies for instance can ensure active participation and fair pricing. However, some additional challenges need to address for fully functional negawatt trading mechanisms in today's energy market.
\end{abstract}
 % ====================================
\section{Introduction}\label{sec:background}Approximately two-thirds of global greenhouse gas emission stems from the energy sector~\cite{Nature_Energy_Egli_2018}. This creates a significant opportunity for prosumers -- participants in the energy grid who both consume and produce energy -- who can buy renewable energy when needed; sell any excess to other consumers; and trade their \emph{right to buy energy} to other prosumers in the network. Such actions allow prosumers to contribute to combating global climate change while saving money. The first two phenomena of managing energy are well established and reported in many existing studies in the literature. For example, see \cite{Lee_NatureEnergy_2019} and \cite{Thomas_Nature_2018}. However, trading the right to buy energy -- also known as negawatt trading~\cite{Jing_Negawatts_GTD_2019} -- is something that is yet to be implemented.

The concept of negawatt was first introduced in the mid-eighties~\cite{Lovins_Mar_1985,Lovins_Keynote_1989} as a technique of energy management. It can be defined as an energy customer's right to buy energy, which is produced due to a change in their energy consumption behaviour. For example, an energy customer may choose to either reschedule their energy related activities to another time or decide to not use energy for some activities and sell their right to buy the saved energy to other entities within the energy network. Note that different demand response programs such as direct load controls, demand bidding programs, demand buyback programs, emergency demand response programs, capacity market programs, and interruptible programs~\cite{ Krarti_Chapter4_2018} also involve reduction in demand from energy consumers. However, the rules and decisions of these programs are set by the grid with very limited inputs from customers. Such incapacity to contribute to pricing decisions has long been a barrier to customers reaping their expected reward from their provision of flexibility to the network. This is due to rules around minimum bid size and difficulties for aggregators to get market access. In contrast, negawatt trading provides prosumers with flexibility and independence in deciding when to reduce their demands, at which price they can sell their right to buy energy, and with whom they want to trade their right of buying, i.e., whether with the grid in the national market or with other prosumers within the network through a peer-to-peer platform. Thus, prosumers can reap their expected rewards from their provision of flexibility to the network by participating in negawatt trading. 

The concept of negawatt trading is also fundamentally different from peer-to-peer energy trading~\cite{Tushar_TSG_2020_Overview}. Peer-to-peer trading is a method of energy trading between two parties within an electricity network without the need for any intermediate entity with possible detrimental impacts on the network if strict constraints on the physical transfer of energy are not maintained~\cite{Chapman_TSG_2018,Baroche_TPWRS_July_2019}. Negawatt trading, on the other hand, is a specific trading mechanism for the trading of the right to buy energy instead of any physical exchange of energy. Therefore, negawatt trading has no detrimental impacts on the electricity network. Further, it can be traded either with the grid with the involvement of intermediate entities or with other customers within a network through a peer-to-peer trading platform without any intermediate entity. 

Thus, negawatt trading offers an unprecedented opportunity to prosumers to participate in the energy market by trading their right to buy energy as opposed to energy demand. However, while advances in the information and communication technology industry have made extraordinary progress in developing energy-efficient devices~\cite{Rogers_Report_2017}, pricing mechanisms~\cite{XingYan_RSER_2018}, building energy management~\cite{Hannan_Access_2018}, local energy trading~\cite{Tushar_TSG_2020_Overview}, energy market structure~\cite{Sausa_RSER_Apr_2019}, and energy policy~\cite{Annunziata_EP_June_2014,Blanchet_EP_Mar_2015}, negawatt trading has remained elusive -- a concept without any serious development. Key reasons have been attributed to the lack of innovation in necessary technologies and a limited understanding of the social, economic and environmental implication of trading the right to buy energy.

Nevertheless, in light of recent developments negawatt trading is starting to be seen as a reality. For example, in Japan, a number of industries have participated in the Yokohama Smart City Project to demonstrate the effectiveness of negawatt hour trading in the future~\cite{Honda_CIRED_2017}. Similar initiatives for demonstrating the feasibility and conceptual architecture of negawatt trading are also being taken in the USA~\cite{Jing_Negawatts_GTD_2019} and Australia~\cite{DMIS,currie2019}. These demonstration projects clearly establish that the technology era with the capacity to enable feasible operation of negawatt trading is finally here.

Considering these recent technological developments, the focus of this study is to provide a multi-disciplinary perspective on the feasibility of negawatt trading in energy markets, which will ultimately contribute to improving the energy efficiency of the smart grid. Note that energy efficiency is not a part of wholesale markets and, at present, there is no mechanism to reward energy efficiency directly for providing  cost savings to consumers. However, in some states in the US, energy efficiency is taken into account when undertaking cost-benefit analyses and calculating the appropriate amount of energy efficiency investment by utilities~\cite{jan_rosenow_2020_3634842}. In the rest of the manuscript, first, we discuss the early challenges that have hindered the large-scale adaptation of negawatt trading in the energy market. Second, we provide a well-grounded evaluation of the recent developments in technologies, which are critically contributing to the feasibility of negawatt trading. Finally, we present a number of outstanding challenges for future research that need to be addressed to ensure the widespread deployment of negawatt trading, followed by some concluding remarks.

\section{Challenges for Wide-Scale Adaptation}\label{sec:challenges}To enable trading negawatt, a typical energy system should have a number of elements including active participants, appropriate communication facilities, secure information systems, suitable market mechanisms, and fair pricing techniques. The main limitations in the capabilities of of each of these network elements that have prevented negawatt trading from wide-scale implementation are summarised in the following.
\subsection{Active participants} In a network, negawatt needs to be exchanged or traded between different prosumers, that is, negawatt sellers and negawatt buyers. Negawatt sellers are the house or building owners who are willing to reduce their energy consumption to trade their rights of buying energy with negawatt buyers. Thus, clearly, for a successful negawatt trading market, prosumers need to actively participate in the exchange. Unfortunately, until very recently, the participation of prosumers has been insufficient due to a number of reasons as discussed below.

Inadequate technical facilities - For trading negawatt, each participating building or house must have enough flexible load, the capacity to monitor energy demand and supply in real-time, capability to interact with its appliances as well as with other buildings, and the ability to make intelligent decisions based on real-time and historic information without compromising on convenience. Unfortunately, a vast majority of buildings in the world today are equipped with inefficient appliances and have no, or very limited, building controls. Consequently, lights are turned on and off manually and air conditioning systems are controlled with local room-to-room switches and thermostats. Building management system solutions are also highly expensive and, thus, difficult to justify for use in small- and medium-size buildings~\cite{Tushar_SPM_Sept_2018}.

Limited prosumer-focused management system - To enable prosumers to create enough negawatt relies on the capabilities of energy management system of the building. Most existing energy management systems are technology-focused, rather than prosumer-focused. However, it is critical that the technology integrates with the users' experience and positively affects prosumers' energy usage behaviour without creating any notable inconvenience~\cite{Tushar_Energy_Apr_2020}. Due to different type of prosumers and their diverse preferences, it is difficult to develop one single energy management platform that captures all prosumers' preference within a network.

Increased concern for privacy and security - Smart energy meters are necessary for most environmental and energy-saving initiatives, including trading negawatt. However,  smart meters can also be used as surveillance tools~\cite{Wang_TSG_May_2019}. For example, the data collected by a smart meter can easily reveal how many showers the occupants in a house have had, when they are cooking, and when they are in and out of the home. Thus, concerns for privacy and security arising from the use of such smart energy solutions has been a critical issue~\cite{Zhang_ICM_Jan_2017} that prevents prosumers from participating in negawatt or other energy exchange programs.

Lack of education - Education can have a profound impact in changing people's behaviour in combatting climate change~\cite{Alison_Climate_Change_2012} and their energy use~\cite{Dowd_EP_Dec_2012}. For example, by motivating people to use renewable energy, adopting a lifestyle that helps  reducing energy demand, and share renewable energy with one another. Nevertheless, environmental education has not been emphasised in either secondary or post-secondary studies. Hence, most prosumers are not motivated to change their energy usage behaviours, which has been a barrier for adopting environmental solutions like negawatt trading in the energy market. Similarly, wider communication and education about the likely financial benefits of trading energy by household prosumers has been largely absent from the discussion.
\begin{figure}[t]
\centering
\includegraphics[width=\columnwidth]{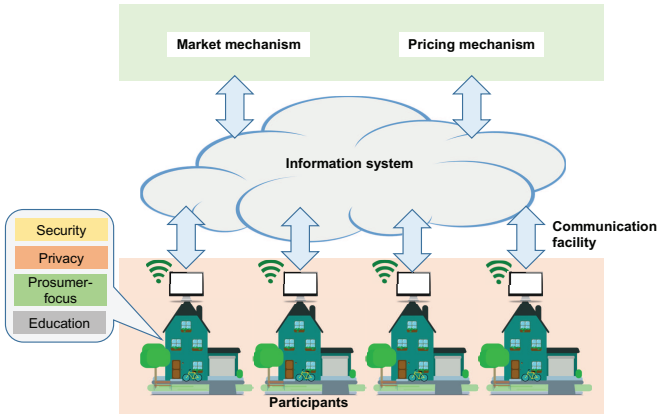}
\caption{\textbf{Challenges of negawatt trading-}~A number of prosumers, such as grid-efficient buildings, are located in a community that want to participate in negawatt trading. Now, to enable this buildings to trade negawatt among themselves, a number of challenges at different aspects of the negawatt trading framework need to address. Prosumers need to actively involved in the negawatt trading, for which educating them about the benefit of such participation is essential. Prosumers also need to be socially interactive to reap the maximum benefit of negawatt trading. There should be appropriate  communication facilities for prosumers within the energy network to securely communicate with one another without compromising their privacies. Further, all information exchange and monetary transactions should be done through a secured information system. All participants should have equal access to market mechanism and the price per negawatt that each participants will receive should be decided in a fair manner. Finally, the regulation of the region should environmentally friendly and support the establishment of negawatt trading to integrate negawatt trading into the existing market and supply system. (Source of clip art within the figure: https://pixabay.com/ )}
\label{Fig:Challenges}
\end{figure}
\subsection{Social engagement}Social engagement and relationships are important in shaping individual and group decisions and actions about energy consumption. Key elements include social and personal norms~\cite{Schwartz_JPSP_1968}, social identity~\cite{Tajfel_PIR_1986} and trust~\cite{Siegrist_RA_Apr_2000}. We know that when individuals are faced with making decisions around energy consumption they will be influenced by expectations of their peers about what is acceptable, the likely consequences of their choices as well as their trust in the actors and institutions involved~\cite{Huijts_RSER_Jan_2012}.  For example, when people gather at social events, they may discuss their use of energy and why they consume in the ways they do and weigh up whether there is any motivation or peer pressure to change or adopt new behaviours~\cite{Dowd_EP_Dec_2012}. \cite{Hargreaves_NatureEnergy_Mar_2020} also stressed the importance of social relationships in shaping energy behaviours including those with friends and family, agencies and community groups and who they identify with. However, the impact of social relations on energy use has often been neglected in the design of energy markets. Failing to cultivate relationships or understanding the social motivations that exist across energy consumers can be detrimental for enabling negawatt sharing between prosumers.
\subsection{Appropriate communication facilities}For trading negawatt, prosumers need to communicate with one another as well as different  energy stakeholders within the network. Therefore, a suitable communication facility is a major requirement for negawatt trading. While there are many communication infrastructures that have been reported in the literature including structured, unstructured, and hybrid architectures~\cite{Jogunola_Energies_Dec_2018}, the choice of a communication architecture that fulfils the performance requirements of latency, throughput, reliability, and security for negawatt trading is critical.
\subsection{Secured information system}A critical element of an energy network, to facilitate negawatt trading between different prosumers, is a well-functioning secure information system. The information system needs to enable all participants to be integrated with the market mechanism, and provide participants with access to the same accurate information about the price, energy status (supply and demand), ancillary market information, and environmental conditions. Importantly, the information system needs to be secured as well as conforming to the privacy requirements of participants.
\subsection{Suitable market mechanism}Market mechanisms consisting of market allocation, payment rules, and clearly outlined pricing formats play a central role in negawatt trading platforms. The main purpose of a market mechanism is to help participants to achieve their desired revenue (or cost reduction) by matching the selling and buying of orders in real-time. Each negawatt seller can influence the maximum availability of negawatt within the market through its demand reduction capacity and subsequent pricing. Different market mechanisms may need to be created and made to co-exist to generate and trade enough negawatt at each stage of market operation, which is yet to be implemented.
\subsection{Fair pricing mechanism}Pricing mechanisms are designed to efficiently balance the energy supply and demand within the network. However, pricing mechanisms for creating and selling negawatt could be different compared to the pricing of traditional energy markets. This is because creating negawatt will not have any marginal cost. Hence, participants may acquire more financial benefits from selling their right to buy energy. However, creating negawatt does involve scheduling and regulating energy usage related activities that may cause additional inconvenience to the negawatt sellers. This inconvenience needs to be taken into account when developing fair pricing mechanisms for negawatt trading. At the same time, the availability of negawatt should influence the set price per unit of negawatt. For example, a higher availability of negawatt within the network may lower the trading price and vice versa.
\subsection{Environmental friendly regulation}The decision of the trading of negawatt in the future electricity market will most likely be governed by regulation and energy policy. Thus, the legislation in a country governs what kind of market design will be allowed, whether there will be any taxes or fees for such trading, and how the negawatt market will be integrated into the existing energy market and supply systems. Governments could provide new policies for system operators, network companies, and utilities to support negawatt trading. For example, distributed system operators could be motivated by new revenue streams and incentives to engage in negawatt trading. Network companies could consider negawatt solutions alongside supply-side options (network upgrades, for example), and the utility could earn a portion of the savings from negawatt trading within the network~\cite{Rosenow_RAP_2019}.

An overview of challenges in different aspects of a negawatt trading framework is shown in Fig.~\ref{Fig:Challenges}.

\section{Evaluation of Enabling Technological Developments}\label{sec:Technologies}To address these challenges, significant technological innovations and developments have been made to establish negawatt trading -- an inter-disciplinary technology with relevance to energy engineering, IoT, optimisation, economics, and human behaviour -- as a feasible practice in energy markets. What follows is an overview of technologies that have laid the pathway for negawatt trading among prosumers and between prosumers and the grid.
\subsection{Grid-interactive efficient buildings}Grid interactive efficient buildings (GEBs)~\cite{Perry_GEB_Oct_2019} are a new concept, in which buildings have the capacity to monitor and control their real-time energy generation and dispatch as well as optimise their energy usage for service, occupant needs and preferences, and cost reduction (or, revenue maximisation) in an integrated fashion~\cite{Neukomm_GEB_Apr_2019}. The main characteristics of a GEB that would enable it to participate in negawatt trading include:

Reliable and low latency communication facility - Each GEB is equipped with equipment that supports two-way connectivity and communication with devices, appliances within the buildings as well as with other GEBs within the network and the grid. Equipment should have the capacity to monitor, report, and provide flexibility to shed, shift, or modulate consumption in response to the control signal sent by the management system of the building.

Intelligent management system - The management system of a GEB can monitor, incorporate, predict, and learn from occupant needs and preferences, outdoor conditions (weather), and from other GEBs' needs and grid requirements. Based on such prediction and learning, it can coordinate and execute complex control strategies that adapt to changing conditions over multiple time scales. Further, the management system can quantitatively estimate and verify the energy and demand savings from different strategies. It can optimise across a choice of multiple strategies to balance efficiency with flexibility and occupancy comfort.

Interoperable, secured, and trusted system - Finally, the overall system responsible for negawatt trading of a GEB should be interoperable and have the capacity to effectively exchange data and control signals among devices, appliances, management systems, and between different GEBs and the grid  in a secure fashion. Data security and protection should be resilient against any cyber attack from unauthorised sources, and as a trusted system, a GEB may need to enforce different specified security policies for performing applications in different contexts.
\subsection{Distributed ledger technology}Distributed ledger technology (DLT) is a digital, shared, and distributed database for recording transactions of assets, and has proven capabilities to improve the efficiency of current energy practices and processes~\cite{Naveed_BlockChain_2019,Lee_NatureEnergy_2019}. In particular, recent advancements in DLT including blockchain, smart contracts, consortium blockchain, Hyperledger, Ethereum, directed acyclic graph, Hashgraph, Holocahin, and Tempo~\cite{Zia_IEEEAccess_Jan_2020} can contribute to negawatt trading due to a number of useful characteristics.

For example, DLT can realise automated billing for prosumers and help negawatt traders with micro-payments, pay-as-you-go solutions, and payment platforms for pre-paid meters. DLT, in conjunction with artificial intelligence techniques, can identify prosumers' energy usage behaviour patterns, and subsequently manage negawatt trading according to individual preferences, energy profiles, and environmental concerns. DLT-enabled distributed trading platforms also have the capacity to be part of market operations such as wholesale market management, commodity trading transactions, and risk management. It enforces trust within the system without the requirement for intermediaries through the use of a distributed ledger, consensus algorithm, and token. Further, DLT has the capacity to improve the control of a decentralised system~\cite{Burger_Report_Nov_2016}. Thus, the adoption of blockchain in the local negawatt trading market can increase behind-the-meter activities such as production of negawatt through regulating self-consumption. Furthermore, DLT can be used for communication of smart devices, data transmission or storage~\cite{Burger_Report_Nov_2016}, which is very important to enable negawatt trading. It can also assist network management of decentralised networks via flexible asset and service management. Finally, DLT has the capability to protect privacy, data confidentiality \cite{Burger_Report_Nov_2016}, and identity management \cite{Diego_CIRED_2017}.  The immutable records and transparency provided by DLT can significantly improve auditing and regulatory compliance of negawatt trading~\cite{Diego_CIRED_2017}.
\begin{figure}[t]
\centering
\includegraphics[width=\columnwidth]{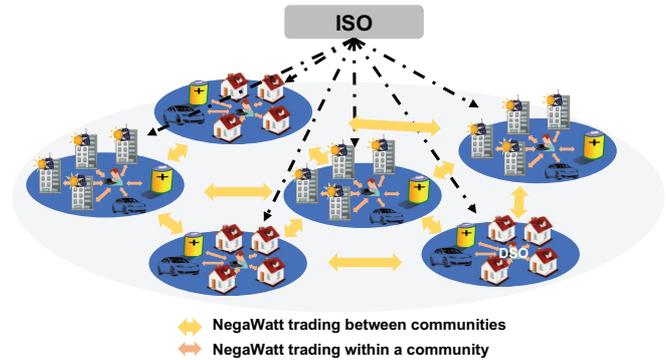}
\caption{\textbf{Market for peer-to-peer negawatt sharing-}~Prosumers can share their negawatt with one another through peer-to-peer sharing. In a community, peers could be apartment buildings, individual homes with distributed energy resources, electric vehicles, and community storage devices. Negawatt can be shared within a community via a community manager such as a distribution system operator (DSO). However, it is also possible to share negawatt among different communities. In such cases, community managers need to coordinate with one another in order to decide on the trading parameters including the price and the amount of negawatt to be traded. A third-party entity such as an independent system operator can facilitate this coordination among different community managers to trade negawatts across communities. (Source of clip art within the figure: https://pixabay.com/ )}
\label{Fig:Market}
\end{figure}
\subsection{Peer-to-peer sharing}In Peer-to-peer sharing, the participants of an energy network can share some of their own resources with one another. These shared resources provide the service and content offered by the network and can be accessed by other peers directly, without the intervention of intermediary entities \cite{P2PDefinition_Aug_2001}. Further, in a peer-to-peer network, any entity can be removed or added, if necessary, without the network suffering from any loss of network service.

Over the last few years, a large number of findings have been reported on the advancement of peer-to-peer sharing in the electricity network. These results cover the aspects of energy cost reduction via peer-to-peer sharing~\cite{Si_AE_Dec_2018}, balancing of supply and demand of energy~\cite{Li_AE_Aug_2019}, engaging prosumers in peer-to-peer sharing~\cite{Kirchhoff_AE_June_2019}, developing appropriate pricing mechanisms~\cite{Thomas_TPS_Early_2018}, identifying uncertainties in peer-to-peer sharing~\cite{Moret_PSCC_June_2018}, transaction security~\cite{Noor_AE_Oct_2018}, and demonstration of pilot projects~\cite{Abrishambaf_ESR_Nov_2019}. Learning from these studies has the potential to significantly advance the research and deployment of negawatt trading within energy systems.  For example, negawatt sharing could be thought of as an alternate version of peer-to-peer trading, in which prosumers' willingness to buy energy, i.e., negawatt, is shared among the participants within decentralised markets~\cite{Sorin_TPWRS_Mar_2019}, community-based markets~\cite{Tushar_TSG_May_2016}, and composite markets~\cite{Liu_TSG_Sept_2018},  instead of watts. An example diagram of such a trading paradigm is shown in Fig.~\ref{Fig:Market}.
\subsection{High speed communication}For successful negawatt trading in energy markets, a critical requirement is a high-speed communication service that is capable of enabling immersive remote operations and interactions with a physical world with low-latency. The 5th generation communication networks (5G) will be fundamental to fulfilling this requirement with their capacity to support a wide range of highly demanding services and applications, pushing the network capabilities to provide extreme performance benefits. This includes the support of massively interconnected devices, and providing necessary services to enable operations and manipulation of physical objects over distance with reliability and low latency, described as the tactile internet in~\cite{Sachs_ProceedingIEEE_Feb_2019}.

Essentially, the 5G network will provide several interdependent tools that will be able to offer the flexibility to enable negawatt trading. For example, 5G networks will be highly programmable and built on network function virtualisation~\cite{Sachs_ProceedingIEEE_Feb_2019}. Thus, more of the network functionality could be implemented in software executed in virtual environments on general purpose hardware, and less specialised hardware built and optimised specifically for certain network functions. A key enabler for a flexible and programmable 5G platforms is the distributed cloud. As a result, the software can be deployed and executed at an optimal place within the network. The software can then become a virtual network function (VNF) or an end-user application software, e.g., BMS of a building, with bi-directional communication for negawatt trading purposes.
\subsection{Internet of Things}Internet of Things (IoT) is essentially the interconnection between computing devices embedded in our everyday life via the Internet, which enables them to send and receive data. It consists of multiple layers including the device layer, network layer, cloud management layer, and application layer. While the device layer is responsible for sensing the environment, collecting data, and controlling flexible loads within a building, the purpose of the network layer is to connect the devices to the application layer. The application layer provides services to the end-users by controlling flexible loads. Examples of such services include demand management, dynamic pricing, energy management, and home security services. The cloud management layer ensures user authentication, along with user and data management. More details of IoT can be found in \cite{Sanjana_WCM_Oct_2016}.

\begin{figure*}[t]
\centering
\includegraphics[width=\linewidth]{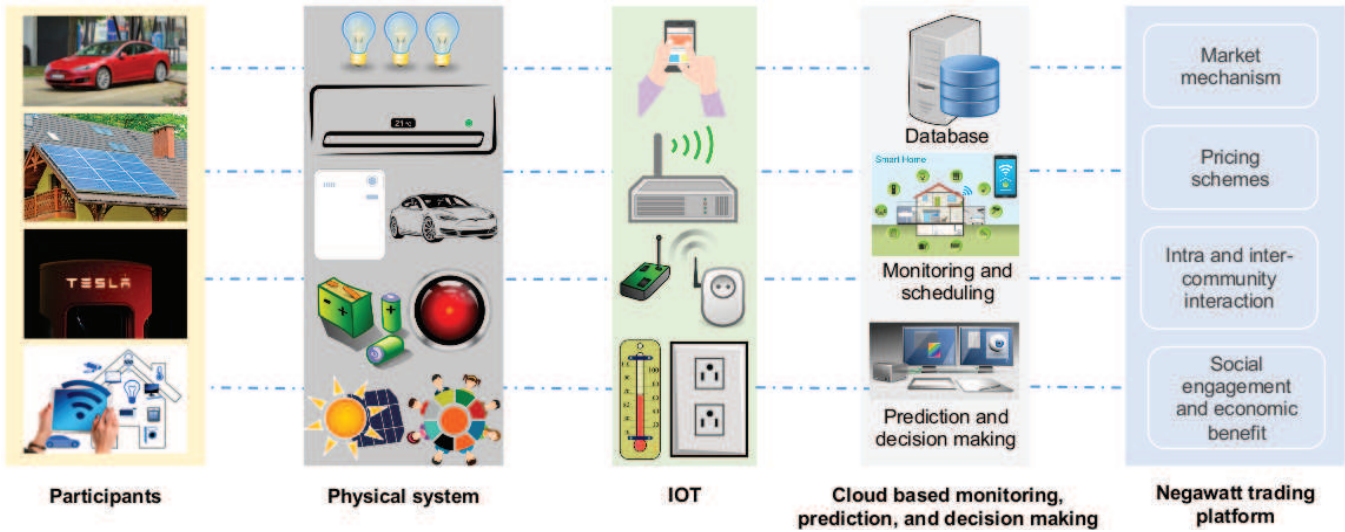}
\caption{\textbf{IoT for prosumers decision-making-}~Internet of Things (IoT) facilitates interconnection between the physical systems of the prosumers via the Internet and enables them to send and receive data over the virtual cloud-based system. Examples of such physical systems include household appliances and distributed energy resources. By converting the physical system to its digital-twin, IoT leverages the negawatt trading platform to sense the environment, collect data, monitor energy usage and generation, and predict and control the use of flexible loads within a building. Thus, IoT enables prosumers to opportunistically participate in both intra-community and inter-community negawatt sharing with appropriate market mechanism and pricing schemes. (Source of clip art within the figure: https://pixabay.com/ )}
\label{Fig:IOT}
\end{figure*}

For participating in negawatt trading, it is important for buildings to be aware -- in real-time -- of energy production, demand, and possibilities of scheduling or throttling of flexible devices through sophisticated sensing and control capabilities. Recent advancements in integrated data acquisition and control systems based on open architectures and cloud-enabled IoT allow building owners (or managers) to monitor and sense buildings' environmental parameters,  collect relevant human activity information, estimate energy usage, and then based on the available energy supply direct BMSs to manipulate the activities of flexible loads according to expectations and specified rules~\cite{Tushar_SPM_Sept_2018}. An overview of application of IoT for negawatt trading is demonstrated in Fig.~\ref{Fig:IOT}.

\subsection{Distributed energy resources}Distributed energy resources (DERs) include behind-the-meter generation, energy storage, inverters, electric vehicles, and controllable loads and their associated applications. DERs offer significant opportunities to reshape the energy future and negawatt trading is one of them. Recent advancement in DERs in terms of their manufacturing material, size of the resources, artificial intelligence, and computational speed has significantly increased their affordability via cost reduction, reliability of operation, and security and privacy. For example, the price of storage has fallen $85\%$ from 2010 to 2018 and is projected to reduce by an additional $18\%$ by 2030~\cite{Logan_Battery_Mar_2019}. Further, with the integration of blockchain with DERs, security and privacy issues have been significantly addressed.

Now, with artificial intelligence and novel computational approaches, participants can simply define rules in their applications, e.g., through their mobile phones, and the transactive meters within their houses can automatically perform trading on their behalf~\cite{Tushar_SPM_July_2018}. Such advancements make it easy for prosumers to trade their negawatt with one another and with the grid and contribute to improving environmental sustainability.
\subsection{Social media platforms}Gaining public engagement in the energy sector has been identified as a major precondition for organisations to achieve sustainability targets. This is due to the fact that sharing knowledge and awareness amongst energy users can have a significant impact on perceptions, beliefs, and attitudes towards the energy transition~\cite{Kimberly_NatureEnergy_2020}. Of course, to enhance engagement, prosumers need to communicate with one another. This has become possible, thanks to the increasing number of interactive social media platforms that are currently available to prosumers as engagement tools. Popular social media platforms like Facebook and Twitter or platforms that are built and managed by energy service providers such as Enosi (https://enosi.io/) or Powerledger (https://www.powerledger.io/)  can significantly help promote negawatt trading across the energy market. They do this by gathering and processing consumers' posts about their energy usage effectively, reducing risks related to misinformation, building strong relationships between prosumers, and engaging a greater number of prosumers in trading by encouraging proactive behaviour.

\subsection{Behavioural economics}Negawatts are a result of reducing demand. While prosumers are gaining greater awareness of the value and need for sustainable energy practices like negawatts, many prosumers still fail to take noticeable steps towards energy efficiency measures due to a significant discrepancy between peoples' self-reported knowledge, values, attitudes and intentions, and their observable behaviour~\cite{Frederiks_RSER_Jan_2015}. With the emergence of the application of behavioural economics in the area of smart grid and energy management~\cite{Saad_Behaviour_2016}, it is becoming possible to reduce such `knowledge-action' and `value-action' gaps. For example, feedback interventions using digital technologies can be very effective at promoting energy conservation behaviour. In particular, real-time feedback on specific and energy-intensive information such as energy usage, negawatt demand in the network, energy price, and potential revenue for negawatt selling may induce considerable behavioural change and production of negawatts~\cite{Tiefenbeck_NE_Nov_2018}. Further, energy education and behavioural programs on the potential impact of negawatt production on energy savings and environmental sustainability could also impact people's energy savings behaviour in the long run. A trial conducted in \cite{Boudet_NE_July_2016} has shown that youth education can potentially influence environmental and/or knowledge, attitudes, and behaviour towards energy demand reduction, and that children's energy behaviour also affects that of parents. Further, prosumers' perceptions of savings, e.g., to generate negawatts, are affected by cognitive processes such as the recall of previous bills~\cite{White_NE_Dec_2018}, and, in the case of negawatt trading, recall of previous economic benefits obtained via trading negawatts with other participants in the energy network.
\subsection{Computational approaches}Substantial work has been done in terms of exploiting computation techniques for trading and sharing resources across energy networks. Examples of such computational approaches include game theory, double auction, constrained optimisation, and artificial intelligence.

Game theory - Game theory  is a mathematical tool that analyses the strategic decision making process of a number of players in a competitive situation, in which the action taken by one player depends on and affects the actions of other players~\cite{Basar_Book_1995}. It has been used extensively in recent years to balance energy supply and demand, develop pricing schemes, increase prosumers' engagement, and provide network services. It has significant potential to be utilised for sharing energy resources including both watts and negawatt within an energy network~\cite{Tushar_SPM_July_2018}.

Double auction - Double auction involves a market of a number of buyers and sellers seeking to interact with one another to trade or share their resources~\cite{Saad_SmartGridComm_2011}. Both buyers and sellers need to truthfully report their bids for efficient and sustainable operation of the market. Double auction has been used in the energy market for demand-supply balance, network services, and prosumers engagement~\cite{Tushar_TSG_2020_Overview}.

Constrained optimisation - Resource sharing in the energy market has been heavily captured via various constrained optimisation techniques including linear programming, non-linear programming, mixed-integer linear programming, and alternating direction method of multipliers (ADMM). For instance, applications of these optimisation techniques can be found in storage management~\cite{Long_AE_Sep_2018}, energy management~\cite{Nguyen_AE_Oct_2018}, and scheduling of flexible loads~\cite{Luth_AE_Nov_2018}.

Artificial intelligence - More recently, artificial intelligence techniques have found many applications in learning the energy usage behaviour of flexible loads and subsequently controlling their energy consumption strategies to achieve certain objectives. Examples of such objectives may include demand-response via reinforcement learning~\cite{Jose_AE_Feb_2019}, building management through deep learning~\cite{Konstantakopoulos_AE_Mar_2019}, and energy cost reduction through an artificial neural network-based approaches~\cite{Reynold_AE_May_2018}.

Indeed, the advancement of these computational techniques would also be valuable in capturing the decision-making processes of prosumers, e.g., see \cite{Okawa_TSG_Nov_2017}, in trading their negawatt, while simultaneously optimising their household activities without affecting their preferences. For instance, computational approaches can be utilised for market settlement, pricing mechanisms, forecasting of reduction in demand for individual participants, enabling strategic interaction between different participants (for trading in P2P platforms) and between the retailer and participants (for trading in the retail-based market), and for capturing and quantifying participants' convenience based on each participant's unique circumstances and parameters. Further, by integrating design thinking~\cite{Tushar_Energy_Apr_2020} and motivational techniques~\cite{Tushar_AE_June_2019} with the designed schemes, the computational approaches can be made more prosumer-centric with potential to attract more participants to actively involve in trading negawatt in the energy market.

\section{Remaining Challenges \& Future Research}\label{sec:FutureResearch}Despite recent technological advancement and efforts, a large number of challenges have yet to be addressed before the successful deployment of techniques that can facilitate wide-scale negawatt trading in the energy market can occur. Below is a summary of challenges that need to be addressed for successful implementation of negawatt markets.

\subsection{Appropriate pricing schemes for negawatt trading}\label{challenge:1}Since the production of negawatt relies on managing demand of prosumers, it requires prosumers to either schedule their energy related activities or regulate the energy consumption of the loads. Therefore, there are possibilities that prosumers may experience inconvenience to produce enough negawatt within the network. To compensate such inconvenience and encourage prosumers to participate in negawatt trading markets, suitable pricing schemes are necessary. Further, inconvenience is not quantifiable like energy and can vary extensively for different energy customers due to different circumstances of generation, demand, preferences, and views of environmental sustainability. As a consequence, the same price per unit of negawatt may not reflect the true reward for all negawatt producers to participate in such trading. Therefore, appropriate price discrimination will need to be introduced in order to ensure a fair and incentive-compatible revenue for all negawatt traders within the market.

\subsection{Network security with low computational expenses}\label{challenge:2}Two critical factors for negawatt trading are trust and security. While identity check and verification might be necessary for prosumers to participate in trading, the security of data injection needs to be ensured in an inexpensive way. At present, blockchain is being considered as the most appropriate trading platform due to its capability to ensure secured and trusted transactions. However, as detailed in \cite{Fairley_Spectrum_Oct_2017}, providing secured trading transactions via blockchain is very computationally expensive. Thus, adopting such a computationally expensive technique will require extensive power to serve the participants. As a consequence, participants will need to share the cost of this service, which will make the trading of negawatts very expensive. Hence, for the sustainability of the market in terms of cost, security and trust, new and computationally less expensive platforms will be required for the wide-spread realisation of negawatt trading.

\subsection{Comfort and convenience}\label{challenge:3}Since the production of negawatt relies significantly upon the scheduling of activities of flexible loads, there is a high chance that it would affect the comfort and day-to-day activities of building occupants. Hence, algorithms for regulating the schedule of different appliances needs to be data-driven. Further, innovation is required in developing the artificial learning algorithms to take into account the diverse behavioural patterns of users, and subsequently schedule flexible loads within houses such that prosumers' comfort and regular day-to-day activities are not affected. An example of such a model for regulating HVAC's temperature control can be found in \cite{Li_ETC_July_2019}.

\subsection{Data accessibility with privacy}\label{challenge:4}To improve prosumers' decisions about negawatt trading, statistically relevant and accurate energy transaction and usage data need to be made available across communities. However, this accessible data also need to provide privacy to each prosumer. Therefore, demonstrated private data anonymisation needs to be facilitated for the necessary data sharing, while simultaneously providing sufficient accuracy for interrogation of data.

\subsection{A unified framework for both watt and negawatt trading}\label{challenge:5}A suitable peer-to-peer network is necessary for negawatt trading between prosumers. With extensive use of DERs, it is reasonable to expect that a prosumer will be able to participate in both energy and negawatt trading within the energy system. Hence, enough flexibility needs to be ensured within the decision making framework of prosumers so that each prosumer can switch its role as a participant between the negawatt and energy market based on available market information and its existing state of energy. Further, the network should also possess sufficient interoperability to facilitate such switching between markets for a very large number of participants.

\subsection{Defining the roles of different stakeholders}\label{challenge:6}Clearly, different stakeholders would be interested in exploiting negawatt trading to deliver different services to their customers and maintain network security. For example, generators may want to use them for reducing production volatility, distribution network service providers for demand constraint, and retailers may want to combat energy imbalances. Nevertheless, these goals could conflict with one another. Hence, negawatt trading schemes need to be designed such that they do not affect participants' independence and benefits.
\section{Conclusion}\label{sec:Conclusion}While the concept of negawatt trading is not new, this perspective confirms that until recently, there have been a number of barriers which have prevented its deployment. These range from social, technical and economic. However, amidst growing concerns of rising greenhouse gas emissions, energy prices and energy security it seems the time may now be ripe for widespread deployment of negawatt trading.   This has been enabled through improvements in communication and technological advances which were not present in the eighties when the concept was first introduced. In spite of these developments there remain a number of challenges that will need to be addressed to ensure successful implementation of such a scheme. These include establishing a framework that recognises the importance of appropriate pricing schemes, market mechanisms, the importance of trust, security concerns and low barriers to entry. As more successful case study examples are implemented negawatt trading should become a socially accepted norm as part of any energy market platform.
\section*{Acknowledgements}This work was supported in part by the Advance Queensland Industry Research Fellowship AQRF11016-17RD2, in part by the University of Queensland Solar (UQ Solar: solar-energy.uq.edu.au), in part by the SUTD-MIT International Design Centre (idc: idc.sutd.edu.sg), and in part by  the U.S. National Science Foundation under Grants DMS-1736417 and ECCS-1824710.
\section*{Competing interests}The authors declare no competing interests.

%\section*{Data Availability Statement}Due to the nature of publication, this paper neither used any raw data nor any any computer codes during the study. Hence, no computer code or data (other than the list of references) has been made available.
%\bibliographystyle{naturemag}
%\bibliography{NatureEnergy}

\end{document}